\documentclass[twocolumn,showpacs,prl,amsmath,amssymb]{revtex4}
\usepackage{graphicx}
\usepackage{dcolumn}
\usepackage{bm}

\begin{document}
\title{ Electron Transport and Hot Phonons in Carbon Nanotubes }
\author{Michele Lazzeri$^1$}
\author{S. Piscanec$^2$, Francesco Mauri$^1$}
\author{A.C. Ferrari$^2$}
\author{J. Robertson$^2$}
\affiliation{ $^1$Institut de
Min\'eralogie et de Physique des Milieux Condens\'es,
4 Place Jussieu, 75252, Paris cedex 05, France \\
$^2$Cambridge University, Engineering Department,
Trumpington Street, Cambridge CB2 1PZ, UK}
\date{\today}

\begin{abstract}

We demonstrate the key role of phonon occupation in limiting the
high-field ballistic transport in metallic carbon nanotubes.  In
particular, we provide a simple analytic formula for the electron
transport scattering length, that we validate by accurate first
principles calculations on (6,6) and (11,11) nanotubes.  The
comparison of our results with the scattering lengths fitted from
experimental I-V curves indicates the presence of a non-equilibrium
optical phonon heating induced by electron transport.  We predict an
effective temperature for optical phonons of thousands Kelvin.

\end{abstract}

\maketitle

Single wall carbon nanotubes (SWNT) have unique properties that make
them strong candidates for future electronic devices.  They can act as
one-dimensional quantum wires with ballistic electron transport
~\cite{todorov,javey03,tans97}. Due to the strong C-C bond, they can
carry the highest current density of any material before they
break. This makes them the best candidates as interconnects in
integrated circuits and for high performance field effect
transistors. It is thus essential to understand what ultimately limits
the high current performance of nanotube devices. High field transport
measurements have shown that electron-phonon coupling limits the
ballistic behavior~\cite{dekker00, javey04,park04}. At high bias
($\ge$0.2V) electrons scatter with optical phonons, while at low-bias
($<$0.2V) they scatter with acoustic phonons. Ballistic transport is
possible up to few hundreds nanometers in low bias, but the electron
mean free path significantly drops at high bias~\cite{dekker00,
javey04,park04}. By constructing devices smaller than the high bias
scattering length a boost in performance is achieved
\cite{javeyPNAS,seidel05}.

The Boltzmann theory is used in Refs.~\cite{dekker00,javey04,park04}
to fit the scattering length ($l_{\rm op}$) from the measured I-V
curves. It is striking that the analysis of several SWNTs, with 1-3~nm
diameters, gives a similar $l_{\rm op}\sim10-15$~nm for optical
phonons back-scattering~\cite{dekker00,
javey04,park04}. Alternatively, the scattering lengths could be
derived knowing (i) the electron-phonon-coupling (EPC), and (ii) the
phonon occupation during transport. Thus, if the EPC is independently
known, the phonon occupation can be derived from the experimentally
fitted scattering length.  A reliable determination of the EPC is
essential to apply this procedure.

Several attempts to determine the EPC using tight-binding (TB) have
been reported
~\cite{dekker00,park04,saito04,perebeinos05,jiang05,pennington04,mahan03}.
The predicted EPCs are not reliable since they strongly depend on the
used TB parametrization.  E.g., for the graphene unit cell and the
E$_{2g}$ phonon, the maximum of the EPC square at ${\bm \Gamma}$
$|D_{\bm \Gamma}^{max}|^2$, has been calculated to be 17, 42, 35/58,
164 (eV/\AA)$^2$ in Refs.  ~\cite{saito04},
~\cite{perebeinos05,nota01}, ~\cite{mahan03,nota01} and ~\cite{park04}
respectively.  These numbers give an order of magnitude spread in the
calculated scattering length, being this proportional to $|D_{\bm
\Gamma}^{max}|^2$.  On the other hand, the optical phonons EPC were
never directly measured.  But, we recently showed that graphite EPCs
can be directly determined from the experimental phonon-dispersions or
from the Raman D peak dispersion~\cite{piscanec04}. The measured EPCs
are in excellent agreement with accurate density functional theory
(DFT) calculations~\cite{piscanec04}. This confirms the reliability of
DFT for determining the EPCs in graphitic materials, such as SWNTs.

In this Letter, we demonstrate the key role of phonon occupation in
limiting the high-field transport in nanotubes.  To do so, we use DFT
to compute with high accuracy the optical-phonons EPC in (6,6) and
(11,11) SWNTs.  We demonstrate that curvature does not affect the EPCs
in SWNTs with diameters $\gtrsim$0.8~nm, such as those used in
transport
measurements~\cite{dekker00,javey04,park04,javeyPNAS,seidel05}.  Then,
we prove that the EPCs for SWNTs of arbitrary chirality can be derived
from those of graphite, by using a simple analytic formula.  We use
our computed and measured~\cite{piscanec04} EPCs to obtain the
electron mean free path for optical phonon scattering in high-field
quasiballistic transport. We obtain the phonon occupation by comparing
our scattering length with the $l_{\rm op}$ fitted in transport
measurements.


SWNTs are identified by the chiral indexes (n,m), giving the chiral
vector ${\bf C_h}=n{\bf a_1}+m{\bf a_2}$, with ${\bf a_1}$ and ${\bf
a_2}$ the graphene lattice vectors~\cite{fig1}. A SWNT is obtained by
folding graphene so that that the two atoms connected by ${\bf C_h}$
coincide. In real space, a SWNT is periodic along its axis according
to the translational vector ${\bf T}$ ~\cite{fig1}. For diameters
$\gtrsim$0.8 nm the electron states are very well described by the
folded graphene bands~\cite{zolyomi04}. Thus, we can safely use
zone-folding to describe the states relevant to charge transport. The
SWNT electron-states correspond to the graphene states having
periodicity ${\bf C_h}$, {\it i.e.}  such that ${\bf k} \cdot {\bf
C_h}= 2 \pi o $, $o$ being an integer.  The SWNT electron states are
labeled by two-dimensional momentum vectors ${\rm \bf k}$, which cut
with a series of parallel lines the graphene Brillouin zone (BZ)
~\cite{fig1}.

The SWNT optical phonon frequencies are modified by the quantum
confinement of electron states along the
circumference~\cite{Dubay,piscanec04,piscanec05}.  For
diameters~$\gtrsim$0.8 nm, at room temperature, the deviation of the
SWNT phonon-frequencies from folded graphene is $\leq$5\%
~\cite{piscanec05}.  Such deviation has a negligible impact on
scattering lengths. Thus, we use zone-folding to describe SWNT phonon
frequencies and eigenmodes, and we label the phonons by a
2-dimensional momentum vector $\bf q$ in the graphene BZ. The SWNT
electron and phonon states are normalized in the 1-dimensional SWNT
unit cell, with period ${\bf T}$.

The decay time for an electron state $\bf k$, band $i$ and energy
$\epsilon_{{\bf \bf k}i}$ into the electron band $j$ and energy
$\epsilon_{({\bf \bf k+q})j}$, due to scattering with $\bf q$~or~$\bf
-q$ phonons in branch $\eta$, and with energy $\hbar \omega_{{\bf
q}\eta}$, is given by the Fermi golden rule~\cite{nota02}:
\begin{eqnarray}
\frac{1}{\tau} &=& \frac{2\pi}{\hbar N} \sum_{{\bf q}\eta} |g_{({ \bf
k}+{ \bf q})j,{ \bf k}i}|^2 \left\{ \delta[\epsilon_{{ \bf k}i} -
\epsilon_{({ \bf k}+{ \bf q})j} + \hbar\omega_{ {\bf q}\eta}] n_{ {\bf
q}\eta} + \right. \nonumber \\ && \left.  \delta[\epsilon_{{ \bf k}i}
- \epsilon_{({ \bf k}+{ \bf q})j} -\hbar\omega_{ {\bf q}\eta}] (n_{
{\bf -q}\eta}+1) \right\}
\label{eq1}
\end{eqnarray}
where $N$ is the number of q points, the two Dirac $\delta$
distributions describe the processes of phonon absorption and
emission, and $n_{{\bf q}\eta}$ is the phonon occupation number.  For
thermal equilibrium, $n_{ {\bf q}\eta}$ is the Bose-Einstein
occupation factor $n_{{\bf q}\eta}=[exp(\hbar\omega_{{\bf
q}\eta}/k_BT)-1]^{-1}$. Within DFT, the EPC is defined as $g_{({ \bf
k}+{ \bf q})j,{ \bf k}i} = D_{({ \bf k}+{ \bf q})j,{ \bf k}i}
\sqrt{\hbar/(2M\omega_{{\bf q}\eta})},$ where $M$ is the atomic mass,
\begin{equation}
D_{({\bf k}+{\bf q})j,{\bf k}i}= \langle {\bf k}+{\bf q},j| \Delta
V_{{\bf q\eta}} |{\bf k},i\rangle,
\label{eq2}
\end{equation}
$|{{\bf k},i}\rangle$ is the electron state, and $\Delta V_{{\bf
q}\eta}$ is the Kohn-Sham potential derivative with respect to phonon
displacement.

For optical phonons, the phonon dispersion is much smaller than the
electron band dispersion. Thus, considering only phonon emission,
i.e. only the second $\delta$, Eq.~\ref{eq1} becomes
\begin{equation}
\frac{1}{\tau} = \sum_{\eta} \frac{\pi}{M\omega_{{\bf q}\eta}} |D_{({
\bf k}+{ \bf q})j,{ \bf k}i}|^2 \rho[\epsilon_{({ \bf k}+{ \bf
q})j}]~(n_{ {\bf -q}\eta}+1),
\label{eq3}
\end{equation}
where $\rho[\epsilon_{({ \bf k}+{ \bf q})j}]$ is the electron density
of states for band $j$, and ${ \bf q}$ is fixed by $\epsilon_{({ \bf
k}+{ \bf q})j} = \epsilon_{{ \bf k}i}-\hbar\omega_{ {\bf q}\eta}$.

\begin{figure}
\centerline{\includegraphics[width=85mm]{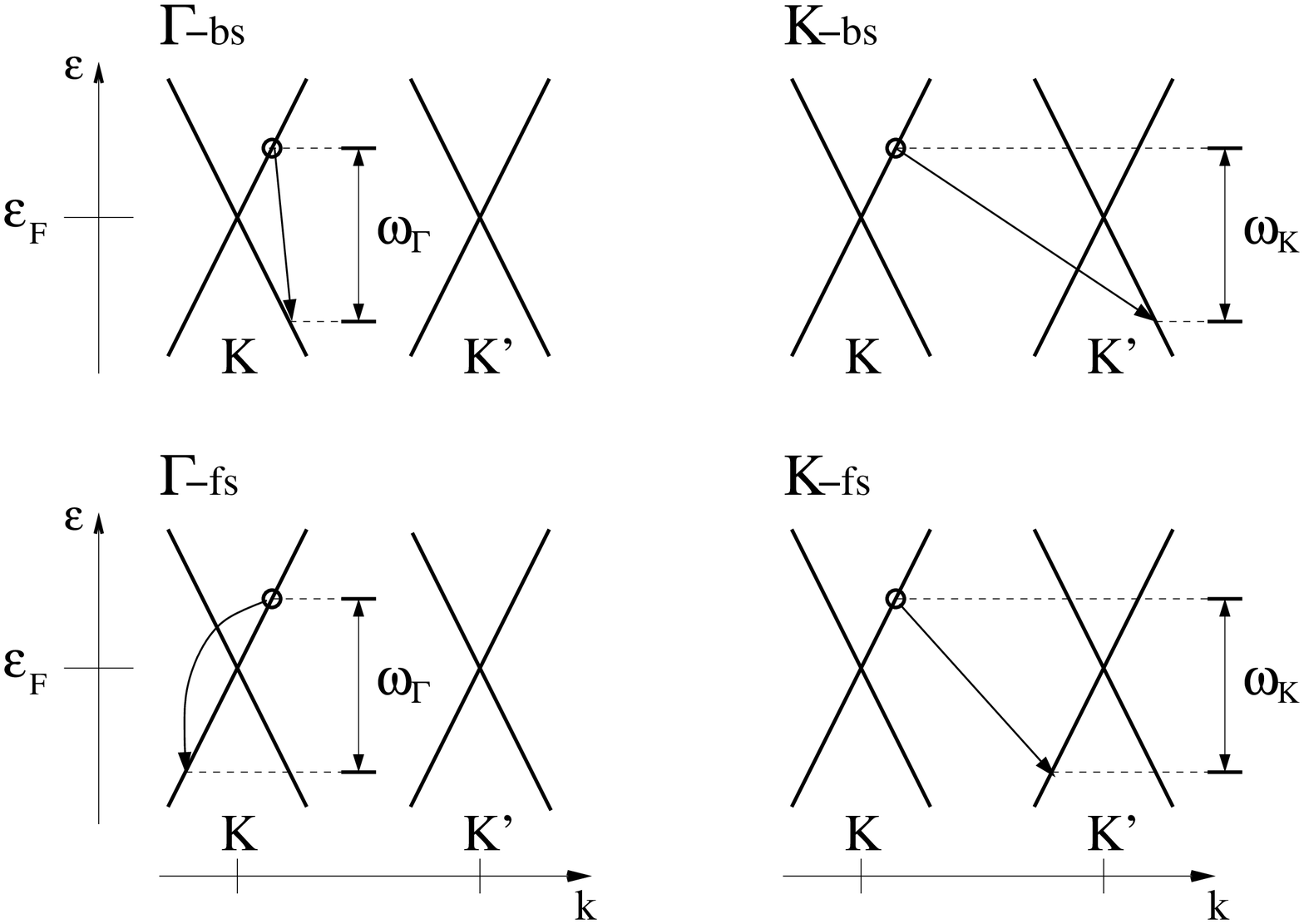}}
\caption{Conduction band electrons in a metallic SWNT.  First order
decay induced by the emission of a phonon with frequency $\omega_{\bf
\rm q}$. An electron state can be scattered into the same band by
forward scattering (fs) or into another band by back scattering (bs).
The resulting four possible processes are labeled by the phonon vector
(${\bm \Gamma}$ or ${\bf K}$) and by the type of scattering (fs or
bs). $\epsilon_F$ is the Fermi energy.} \label{fig2}
\end{figure}

For graphene, the electronic gap is zero for the $\pi$ bands at the
two equivalent BZ points ${\bf K}$ and ${\bf K'}=2{\bf K}$.  Metallic
SWNTs have electron states corresponding to the graphene ${\bf K}$ and
${2\bf K}$ points. For bias smaller than 1.0 V, the mobile electrons
belong to $\pi$ bands near the Fermi energy $\epsilon_F$. Both the
initial and final states, ${ \bf k}$ and ${ \bf k}+{\bf q}$, are close
to ${ \bf K}$ or to ${ 2 \bf K}$. The $\pi$ bands can be considered
linearly dispersive, crossing at $\epsilon_F$ with slopes $\beta$ and
$-\beta$, with $\beta=5.52$~\AA$\cdot$eV, and density of states
$\rho[\epsilon_{({ \bf k}+{ \bf q})j}]=T/(2\pi \beta)$. Phonons
connect electron states near $\epsilon_F$ only for ${\bf
q}\simeq{\Gamma}$ of for ${\bf q}\simeq{\bf K}$, thus, in
Eq.~(\ref{eq3}), $\omega_{{\bf q\eta}}$ can be approximated by
$\omega_{{\bf \Gamma}\eta}$ or by $\omega_{{\bf K}\eta}$. After the
scattering, the electron can either maintain its propagation
direction, forward scattering (fs), or reverse it, back scattering
(bs). We thus have four possible scattering processes,
Fig.~\ref{fig2}. We label the corresponding decay times $\tau^{\rm
bs}_{\bf \Gamma}$, $\tau^{\rm bs}_{\bf K}$, $\tau^{\rm fs}_{\bf
\Gamma}$, and $\tau^{\rm fs}_{\bf K}$. The total decay time is given
by $1/\tau=1/\tau^{\rm bs}_{\bf \Gamma}+ 1/\tau^{\rm bs}_{\bf K}+
1/\tau^{\rm fs}_{\bf \Gamma}+1/\tau^{\rm fs}_{\bf K}$. To model the
transport with Boltzmann~\cite{dekker00,javey04} one needs to know the
decay times for each individual process.  Possible scattering into
higher energy bands is neglected. This assumption is valid for
electron energies lower than $2\beta/d$, $d$ being the tube
diameter. Note that the electron-phonon scattering processes involved
in electron transport are entirely analogous to those involved in
Raman spectroscopy~\cite{piscanec04,Thomsen00,Sait}.

We now compute the SWNT EPCs by folding the corresponding graphene
EPCs. We then demonstrate the validity of this approach, which, so
far, has never been proven. The EPC ($D$) for a SWNT of arbitrary
chirality can be obtained from the graphene EPC ($\tilde D$), using:
\begin{equation}
S~|D_{({ \bf k}+{ \bf q})i,{ \bf k}j}|^2= \tilde S~|\tilde D_{({ \bf
k}+{ \bf q})i,{ \bf k}j}|^2,
\label{eq5}
\end{equation}
where $\tilde S = a_0^2 \sqrt{3}/2 $ is the graphene unit-cell
surface, $a_0=2.46$~\AA~is the graphene lattice spacing, and
$S$=${|{\bf C_h}\times{\bf T}|}$ is the nanotube unit-cell surface
~\cite{fig1}.  Eq.~\ref{eq5} comes from Eq.~\ref{eq2}, using the
appropriate normalization of phonon and electron states.  Combining
all the previous results, for metallic SWNTs, Eq.~\ref{eq3} becomes:
\begin{equation}
\frac{1}{\tau} =\frac{1}{d} \sum_\eta\frac{\sqrt{3}a_0^2|\tilde D_{({
\bf k}+{ \bf q})j,{ \bf k}i}|^2} {4 \pi M\omega_{{\bf q}\eta}\beta }
(n_{{\bf -q}\eta}+1),
\label{eq6}
\end{equation}
where $d=C_h/\pi=a_0\sqrt{n^2+nm+m^2}/\pi$ is the diameter.

For electron states near $\epsilon_F$ in graphene, $|\tilde D|^2$ is
not negligible only for the ${\bm \Gamma}-{\rm E}_{2g}$ and ${\bf
K}-{\rm A}'_1$ modes\cite{piscanec04}. For optical phonons near
${\bm\Gamma}$, the doubly degenerate E$_{2g}$ mode splits into two
almost longitudinal (LO) and transverse (TO) modes. For small ${\bf
k'}$, the corresponding EPCs are:
\begin{eqnarray}
|\tilde D_{({\bf K+k'+q})\pi^*,({\bf K+k'})\pi}^{\rm LO/TO}|^2 &=&
\langle \tilde D^2_{\bm \Gamma}\rangle_{\rm F}
~[1\pm\cos(\theta+\theta')] \nonumber \\ |\tilde D_{({\bf
K+k'+q})\pi^*,({\bf K+k'})\pi^*}^{\rm LO/TO}|^2 &=& \langle \tilde
D^2_{\bm \Gamma}\rangle_{\rm F} ~[1\mp\cos(\theta+\theta')],
\label{eq7}
\end{eqnarray}
with $\theta$ the angle between ${\bf k'}$ and ${\bf q}$, $\theta'$
the angle between ${\bf k'}+{\bf q}$ and ${\bf q}$, and $\pi$ and
$\pi^*$ are the lower and higher $\pi$ bands~\cite{piscanec04}. $+$ or
$-$ refer to the LO and the TO modes, respectively.  For the ${\bf
K}-{\rm A}'_1$ mode we have:
\begin{eqnarray}
|\tilde D_{(2{\bf K+k'+q'})\pi^*,({\bf K+k'})\pi}|^2&=& \langle \tilde
D^2_{\bf K}\rangle_{\rm F} ~(1+\cos\theta'') \nonumber \\ |\tilde
D_{(2{\bf K+k'+q'})\pi^*,({\bf K+k'})\pi^*}|^2&=& \langle \tilde
D^2_{\bf K}\rangle_{\rm F} ~(1-\cos\theta''),
\label{eq8}
\end{eqnarray}
where $\theta''$ is the angle between ${\bf k'}$ and ${\bf k'}+{\bf
q'}$~\cite{nota00}.  From our DFT calculations, $\langle \tilde
D^2_{\bm \Gamma}\rangle_{\rm F} = 45.60$~(eV/\AA)$^2$, and $\langle
\tilde D^2_{\bf K}\rangle_{\rm F} = 92.05$~(eV/\AA)$^2$~\cite{nota01}.
These values reproduce very accurately the EPCs extracted from the
measured graphite phonon dispersion~\cite{piscanec04}.

From Eqs.~\ref{eq7}-\ref{eq8} we obtain the EPCs for all the
scattering processes of Fig.~\ref{fig2}. The results are in
Tab.~\ref{tab1}. Note that for metallic SWNTs, ${\bf k'}$ and ${\bf
q}$ or ${\bf q'}$ are always parallel to the axis, and $\theta$,
$\theta'$ and $\theta''$ are always equal to 0 or $\pi$.  Thus, the
results of Tab.~\ref{tab1} do not depend on chirality.
\begin{table}
\begin{ruledtabular}
\begin{tabular}{ccc|ccc}
\multicolumn{3}{c}{Back-scattering}
&\multicolumn{3}{c}{Forward-scattering}\\ ${\bm \Gamma}$-LO&${\bm
\Gamma}$-TO&{\bf K}& ${\bm \Gamma}$-LO&${\bm \Gamma}$-TO&{\bf K}\\
\hline &&&&&\\ $2\langle \tilde D^2_{\bm \Gamma}\rangle_{\rm
F}$&0&$2\langle \tilde D^2_{\bf K}\rangle_{\rm F}$ & 0&$2\langle
\tilde D^2_{\bm \Gamma}\rangle_{\rm F}$&0\\
\end{tabular}
\end{ruledtabular}
\caption{$|\tilde D|^2$ between $\pi$ states around ${\bf K}$.}
\label{tab1}
\end{table}

Using Eq.~\ref{eq6} and Tab.~\ref{tab1} we can easily compute the
scattering length $l$ ($l=\tau\beta/\hbar$) for a generic metallic
SWNT.
\begin{equation}
l_{{\bf q}\eta} = \alpha_{{\bf q}\eta} d / (n_{{\bf-q}\eta}+1);~~~
\alpha_{{\bf q}\eta}=\frac{4\pi}{\sqrt{3}} \frac{M\omega_{{\bf
q}\eta}\beta^2}{\hbar a_0^2 |\tilde D|^2}. \label{eq9}
\end {equation}
In Refs~\cite{dekker00,javey04,park04} phonons are assumed thermalized
at room temperature ($n_{{\bf-q}\eta}\simeq0$) and the contribution of
forward scattering is neglected. Thus, to compare with experiments, we
consider only back scattering and define $l_{\rm op}=(1/l^{\rm
bs}_{\bf K}+1/l^{\rm bs}_{{\bm \Gamma}-{\rm LO}})^{-1}$. Using the
graphene phonon frequencies ($\omega_{\bf \Gamma}=196.0$~meV,
$\omega_{\bf K}=161.2$~meV)~\cite{piscanec04}, we obtain the results
of Tab.~\ref{tab2}. In particular, we get a simple scaling between
scattering length and diameter:
\begin{equation}
l_{\rm op}=65\, d.
\label{eq10}
\end{equation}

To validate these results, we perform an explicit calculation of
$|D|^2$ on (6,6) and (11,11) armchair SWNTs. DFT calculations are done
with the gradient corrected functional of Ref.~\cite{PBE}. We use
plane-waves (40 Ry cutoff) and pseudopotential
approaches~\cite{Pseudo}. Electron states are occupied with an
Hermite-Gauss smearing of order 1 and a width of 0.05
Ry~\cite{Methfessel}.  EPC calculations are done using the
perturbative method of Ref.~\cite{DFPT}. We use an hexagonal
super-cell with a neighboring tube distance of 5~\AA~and a grid of 16
k-points with a logarithmic distribution around the ${\bf K}$
point. In Tab.~\ref{tab2} we report the scattering lengths obtained
using the $|D|^2$ computed explicitly for the two tubes.
Tab.~\ref{tab2} shows that the zone-folding results are quite accurate
for the 1.5 nm diameter tube, and only a $<$20\% difference is present
for the smaller-diameter tube, due to curvature effects.

\begin{table}
\begin{ruledtabular}
\begin{tabular}{l|ccc}
  & d=0.8 & d=1.49 & d=2.5\\ \hline $l^{\rm bs}_{\bf K}$ & 74~(78) &
            137~(141) & 230\\ \hline $l^{\rm fs}_{{\bf\Gamma}\rm-TO}$
            & 183~(189) & 336~(331) & 564\\ \hline $l^{\rm
            bs}_{{\bf\Gamma}\rm-LO}$ & 183~(215) & 336~(362) & 564\\
            \hline $l_{\rm op}$ & 53 & 97 & 163\\
\end{tabular}
\end{ruledtabular}
\caption{Scattering lengths (in nm) for the processes in first column.
{\it Phonon occupation is assumed thermalized at room
temperature}. All lengths are in nm. In parenthesis, results from
direct DFT calculations on (6,6) and (11,11) SWNTs with diameter
$d=0.8$ and 1.49 nm.  The other values are obtained by folding the
graphene EPCs. } \label{tab2}
\end{table}


Thus, Eq.~\ref{eq9}, combined with our {\it computed and measured}
graphite EPCs~\cite{piscanec04}, provides an independent assessment of
the scattering lengths for the 1-3 nm diameter SWNTs of
Refs.~\cite{dekker00,javey04,park04}.  However, the scattering length
reported in Refs.~\cite{dekker00,javey04,park04} is~$\sim$10-15
nm. This is one order of magnitude smaller than what is predicted by
Eq.~\ref{eq10}.  This disagreement cannot entirely be attributed to
the experimental uncertainty in measuring the scattering lengths,
since three independent experiments on a variety of different tubes
obtained very similar results~\cite{dekker00,javey04,park04}. Also, it
cannot be ascribed to an error in the DFT EPCs, since our computed
EPCs reproduce very well those extracted from the experimental
graphite phonon dispersions or from the Raman D-peak dispersion
~\cite{piscanec04}. Thus, we conclude that the hypothesis of
thermalized phonon occupation ($n_{{\bf-q}\eta}\simeq 0$ in
Eq.~\ref{eq9}) does not hold.

A significant phonon occupation $n$ can explain the small value of the
measured scattering length $l_{\rm op}$. This hot phonon generation
can occur if, during high-field electron transport, the optical
phonons excitation rate is faster than their thermalization rate.
With a high phonon occupation, both phonon emission and absorption
processes are equally relevant. We can estimate the scattering lengths
by assuming that the phonon occupation $n$ is independent of $\bf q$
and $\eta$, and by summing the absorption and emission
contributions. The scattering lengths are then obtained by
substituting $(n_{{\bf-q}\eta}+1)$ with $(2n+1)$ in Eq.~\ref{eq9}:
\begin{equation}
l_{\rm op}=65~d/(2n+1).
\end{equation}
$n$ in the 2.7-5 range is necessary to reconcile the scattering
lengths derived from the computed and measured EPCs and form the fit
of the measured IV curves. This corresponds to an effective
temperature for the occupation of optical modes $\gtrsim$ 6000 K. This
temperature is related only to the phonons directly excited by first
order scattering, and not to other phonons, otherwise the SWNT would
melt. Thus, during high-bias electron transport, the phonons are not
in thermal equilibrium.

These results are consistent with the observation that high-bias
saturation currents in SWNTs on a substrate are significantly higher
than those in suspended SWNTs~\cite{cao05}.  Indeed, the effective
temperature of optical phonons in suspended SWNTs is expected to be
higher due to the absence of a thermally conductive substrate for heat
sinking.  Also, it was recently reported that electrons inelastically
tunnelling from a scanning tunnelling electron microscope tip into a
SWNT induce a non-equilibrium occupation of the radial breathing mode
phonons~\cite{nature2004}.  In this case, the tunnelling electrons are
mostly coupled with the radial breathing mode, since the associated
vibrations directly modulate the tip-nanotube distance.  On the
contrary, in high-field transport experiments, electrons mostly excite
optical phonons, which are the dominant scattering source.

The generation of hot optical phonons, during high field electron
transport, can be directly detected by a Raman scattering experiment
combined with electron transport.  The occupation of a given phonon
mode can be determined from the ratio of the corresponding Stokes and
anti-Stokes peak intensities.  A rise in this ratio would be direct
evidence and would allow to quantify the hot phonons generation.

Finally, our approach can be easily extended to calculate scattering
times and mobility in semiconducting SWNTs.  In this case, the
scattering times depend on electron energy and chirality since the
bands are hyperbolic and the electron density of states is energy
dependent. The angles $\theta$, $\theta'$, and $\theta''$, entering in
Eqs.~(\ref{eq7}) and (\ref{eq8}), are not equal to 0 or $\pi$, as in
metallic SWNTs, but depend on the chirality and on the initial and
final state momenta ($\bf k$ and ${\bf k}+{\bf q}$). These angles can
be computed by simple geometrical considerations.

In conclusion, accurate DFT calculations of EPCs in graphene, (6,6)
and (11,11) SWNTs, combined with a simple zone-folding model and the
experimental graphite EPCs are used to interpret the saturation SWNTs
I-V curves of Refs.~\cite{dekker00,javey04,park04}.  We show that the
optical phonon occupation is greatly increased during high-bias
electron transport, with an effective temperature of thousands K. Such
a high temperature causes a strong reduction of the ballistic
scattering length.  This suggests coupling the optical phonon mode
with an heat sink in order to reduce their effective temperature.
This would increase the scattering length up to the maximum value of
Eq.\ref{eq10}, which sets the ultimate limit of ballistic transport.
Finally, the phonon generation by electron scattering is analogous to
the photon generation by stimulated emission in semiconducting lasers.
This suggests that SWNTs, under high bias, can act as a possible
source of coherent phonons.

Calculations were performed at HPCF (Cambridge) and IDRIS (grant
051202).  S.P. acknowledges funding from Marie Curie
IHP-HPMT-CT-2000-00209 and A.C.F. from the Royal Society. Funding from
EU project CANAPE and EPSRC grant GR/S97613 is acknowledged.


\begin{thebibliography}{99}
\bibitem{nota01}
In a first-neighbors TB
$|D_{\bm \Gamma}^{max}|^2= \eta^2 3/2$, where
$\eta$ is the derivative of the hopping integral with respect to
the C-C distance. We compute the $|D_{\bm \Gamma}^{max}|^2$ from the
$\eta$ values reported in the mentioned Refs.
According to our DFT calculations
$|D_{\bm \Gamma}^{max}|^2$= 91.2 (eV/\AA)$^2$, corresponding to
$\eta$ = 7.8 eV/\AA.
\bibitem{nota00}
The angular dependence of Eqs.~\protect\ref{eq7}-\protect\ref{eq8}
is due to the phase of the electron wavefunctions. It is derived
by TB and reproduces accurately our DFT
calculations~\protect\cite{piscanec04}.
\end{thebibliography}
\end{document}